\documentclass[%
 reprint, 
 superscriptaddress,
 amsmath,amssymb,
 aps,
 prl
floatfix,
longbibliography
]{revtex4-1}

\usepackage{enumitem}
\usepackage{graphicx}

\usepackage{bm}
\usepackage{xcolor}
\usepackage[colorlinks=true,citecolor=blue,linkcolor=magenta]{hyperref}

\usepackage{lipsum}

\begin{document}

\title{Benchmarking Information Scrambling}
\author{Joseph Harris}
\affiliation{Theoretical Division, Los Alamos National Laboratory, Los Alamos, New Mexico 87545, USA}
\affiliation{Department of Applied Mathematics and Theoretical Physics, University of Cambridge, Cambridge, CB3 0WA, UK}
\affiliation{School of Physics and Astronomy, University of Edinburgh, Edinburgh, EH9 3FD, UK}
\author{Bin Yan}
\affiliation{Theoretical Division, Los Alamos National Laboratory, Los Alamos, New Mexico 87545, USA}
\affiliation{Center for Nonlinear Studies, Los Alamos National Laboratory, Los Alamos, New Mexico 87545, USA}
\author{Nikolai A. Sinitsyn}
\affiliation{Theoretical Division, Los Alamos National Laboratory, Los Alamos, New Mexico 87545, USA}

\date{\today}

\begin{abstract}
Information scrambling refers to the rapid spreading of initially localized information over an entire system, via the generation of global entanglement. This effect is usually detected by measuring a temporal decay of the out-of-time order correlators. However, in experiments, decays of these correlators suffer from fake positive signals from various sources, e.g., decoherence due to inevitable couplings to the environment, or errors that cause mismatches between the purported forward and backward evolutions. In this work, we provide a simple and robust approach to single out the effect of genuine scrambling. This allows us to benchmark the scrambling process by quantifying the degree of the scrambling from the noisy backgrounds. We also demonstrate our protocol with simulations on IBM cloud-based quantum computers.
\end{abstract}

\maketitle


In complex many-body systems, initially localized information quickly spreads throughout the entire system---a process known as information scrambling. Though information is ultimately conserved, it gets encoded into global entanglement among many degrees of freedom, and hence becomes inaccessible by local measurement. Information scrambling was originally studied in the context of black hole physics \cite{Hayden2007Black,Kitaev2015,Maldacena2016Bound}, and has since emerged as a field with a wide-ranging impact across different areas in physics, e.g., quantum chaos in many-body systems \cite{Rozenbaum2017Lyapunov,Roberts2017Chaos,Cotler2017Chaos,Lin2018Out,Garciaa2018Chaos,Yan2020Information,Yan2020Recovery}, phase transition \cite{Sahu2019Scrambling,Choi2020Quantum}, and quantum machine learning \cite{Holmes2021Barren}. Considerable effort has also been made in probing this effect in various experimental systems \cite{Landsman2019Verified,Li2017Measuring,Garttner2017Measuring}.

Information scrambling is usually measured by the temporal decay of the out-of-time order correlators (OTOCs) \cite{Kitaev2015,Larkin1969,Swingle2018Unscrambling}, defined as
\begin{equation}
    \langle W^\dag(t)V^\dag W(t)V \rangle,
\end{equation}
where the average is taken over a given quantum state. $W$ and $V$ are local operators, usually considered to act on distinct subsystems. $W(t)$ is the Heisenberg evolution of $W$, which becomes a global operator as scrambling proceeds, causing decay of the OTOCs. 

However, it is difficult to distinguish between scrambling and decoherence \cite{Yoshida2019Disentangling}: the latter causes leakage of the system information to the environment and, in general, induces decay of the OTOC as well. Protocols that measure the OTOC often involve forward and backward evolution of the system, which in practice do not exactly match each other due to operational errors. This can also cause decay of the OTOC signals. Understanding the behavior of scrambling in presence of decoherence and operational errors is an active and ongoing research in the field \cite{Swingle2018Resilience,Knap2018Entanglement,Syzranov2018Out,Zhang2019Information,Gonzalez019Out,Yoshida2019Disentangling,Joshi2020Quantum,Touil2021Information,Dominguez2021Dynamics,Zanardi2021Information}. In this line of effect, the first experiment that provided a positive verification of scrambling was performed with trapped ions \cite{Landsman2019Verified}, using a teleportation-based protocol \cite{Yoshida2019Disentangling}. This approach uses two copies of the system and an entangled input state (Bell state) between the copies, requiring sophisticated engineering of the system and therefore hindering its practical applications. More recently, the field has witnessed an increasing number of studies of information scrambling in various experimental platforms, e.g., superconductors \cite{Zhao2021Probing}, trapped ions \cite{Joshi2020Quantum}, and cloud-based quantum computers \cite{Braumuller2021Probing,Mi2021Information,Geller2021Quantum}. However, it remains a challenge to design a simple and robust protocol for benchmarking the true signals of scrambling from a noisy background. 

In this work, we propose a solution to this task. Our approach is based on a novel quantum butterfly effect \cite{Yan2020Recovery,Sinitsyn2020Quantum} that can unambiguously distinguish scrambling and nonscrambling dynamics. Due to the global entanglement generated by scrambling, information becomes more robust against local perturbations, and hence, when scrambled and disturbed by local perturbations, can still be partially recovered through a reversed unscrambling process. In contrast, decoherence and errors only temper this effect and weaken the signal produced by true scrambling. In the following, we develop the underlying theory, introduce the benchmarking protocol, and apply it to a model problem of fast scrambling in the presence of decoherence.

\begin{figure}[b!]
    \centering
    \includegraphics[width=\columnwidth]{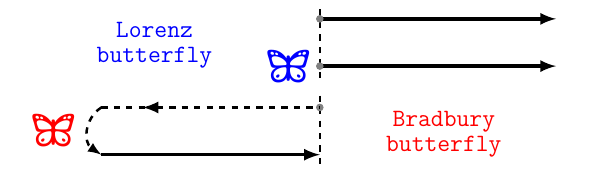}
    \caption{In the Lorenz picture of the butterfly effect, one compares two trajectories evolved under the same Hamiltonian, but with slightly different initial conditions. In the Bradbury picture, perturbation is applied in the past.}
    \label{fig:butterfly}
\end{figure}

In chaotic classical dynamics, small perturbations in the initial conditions can trigger dramatic changes in the time evolution. This effect is well known as the Lorenz butterfly effect. An (classically) equivalent picture of the butterfly effect was introduced by Bradbury \cite{bradbury1952}, where the perturbation is applied in the past (Fig.~\ref{fig:butterfly}).
However, these two pictures exhibit subtle differences in quantum dynamics \cite{Yan2020Recovery,Cao2021Origin}: the overlap between the two trajectories (wave functions) in the Lorenz picture remains a constant during time evolution, due to the unitarity of (isolated) quantum dynamics. On the other hand, in the Bradbury picture, the overlap between the two states at time $t=0$---the initial input state and the final output state after the backward and forward evolution loop---does decay as a function of the evolution time. Moreover, asymptotically, the output state always contains partial information of the initial one, with the amount of information determined by the type of perturbation. This is in sharp contrast with classical chaotic dynamics, which on average smear all the initial information over the entire accessible phase space. For these reasons, we propose to call this phenomenon the quantum antibutterfly effect. 

More precisely, the perturbation in the ``past'' can be described by a general quantum channel $\Lambda$:
\begin{equation}\label{eq:perturbation}
   \rho \rightarrow \Lambda(\rho) = \sum_k M^\dagger_k \rho M_k,\quad \sum_k M_kM_k^\dagger = \mathbb{I},
\end{equation}
where $\mathbb{I}$ is the identity operator, and $M_k$ are the Kraus operators. The initial state $\rho$, after the Bradbury's process aforementioned, becomes
\begin{equation}\label{eq:outstate}
    \rho(t) = U_t\Lambda(U^\dag_t \rho U_t)U^\dag_t,
\end{equation}
where $U_t$ is the evolution operator for a time $t$. This process is also recognized as the quantum twirling channel of the perturbation $\Lambda$ \cite{Dankert2009Exact}. 

After a long time evolution with a chaotic Hamiltonian, when the evolution operator becomes sufficiently random, the asymptotic state can on average be described as
\begin{equation}\label{eq:asymstate}
    \rho_{\rm as} = p\rho + (1-p)\frac{\mathbb{I}}{d},
\end{equation}
where $d$ the dimension of the total Hilbert space and the probability $p$ is determined by the error channel, namely,
\begin{equation}\label{eq:probasym}
p = \frac{\sum_k {\rm Tr} M_k {\rm Tr}M^\dagger_k-1}{d^2 - 1}.
\end{equation}
A detailed derivation of this form is presented in Supplemental Material (SM).

To extract this universal probability $p$, one can measure, e.g., the fidelity of the output state of the twirling channel. Here, we will focus on a similar quantity, the overlap between the final and initial states,
\begin{equation}\label{eq:fidelity}
    F(t)\equiv {\rm Tr}\left[\rho(t)\cdot\rho\right] = \sum_k {\rm Tr} \left[M^\dag_k(t)\rho M_k(t)\rho\right],
\end{equation}
whose asymptotic value has a simple expression
\begin{equation}\label{eq:asymfidelity}
    F_{\rm as} = p{\rm Tr}\left[\rho^2\right] + (1-p)/d.
\end{equation}
One can verify that this asymptotic value of the overlap applies to small subsystems as well, for which the overlap can be evaluated through a state tomography.

We would like to emphasize several remarks:\par
i) The asymptotic state (\ref{eq:asymstate}) is obtained by averaging over an assemble of random unitaries with respect to Haar measure \footnote{Technically, the same result can be obtained for an ensemble of unitaries which is as random as a unitary 2-design, since the average is performed for an expression in terms of the second moment of the unitary.}. However, the fluctuation from the mean is exponentially small in the size of the total system. Consequently, for a single unitary randomly drawn from such an ensemble, deviations from this averaged behavior are exponentially suppressed. \par
ii) The overlap (\ref{eq:fidelity}) is identified as a sum of special types of OTOCs between the state $\rho$ and the Kraus operators. Hence, scrambling causes decay of the overlap in the same manner as to the OTOCs. This particular type of OTOC has been used \cite{Garttner2017Measuring} and demonstrated to have various benefits \cite{Swan2019Unifying}. As will be shown in the following, this quantity can also acquire a large universal asymptotic value (\ref{eq:asymfidelity}), which will respond further to decoherence and errors. This unique feather lies in the core of our protocol for singling out information scrambling.

Inspired by the above theory of the antibutterfly effect, we propose the following protocol to detect and benchmark information scrambling. In this protocol, the total system is sent through a Bradbury process, and only a subsystem is measured:

\begin{enumerate}
    \item[1.] Initialize the total system such that a small subsystem is prepared in state $\rho_S$.
    \item[2.] Evolve the system forward for a time $t$; perturb a different subsystem, and then evolve the system backward for the same time $t$.
    \item[3.] Measure the same subsystem, and evaluate its overlap with respect to the initial state $\rho_S$.
\end{enumerate}

For example, for a (large) spin-$1/2$ system with scrambling dynamics, we prepare a single target spin in any pure state and choose the perturbation channel as a projective measurement on any single spin. The overlap computed with (\ref{eq:asymfidelity}) and (\ref{eq:probasym}) saturates to $\sim 0.75$. On the other hand, in the presence of decoherence and errors, the asymptotic state of the target spin will become random, hence resulting in $\sim 0.5$ overlap. Calculation of these values is presented in the SM.

We also note the similarity between our approach and randomized benchmarking (RB) protocol \cite{Emerson2005Scalable,Magesan2011Scalable} for quantum computers. Indeed, our approach inherits the main benefits of RB, i.e., it is scalable and is independent with errors in state initialization and readout.


\begin{figure}[t!]
    \centering
    \includegraphics[width=\columnwidth]{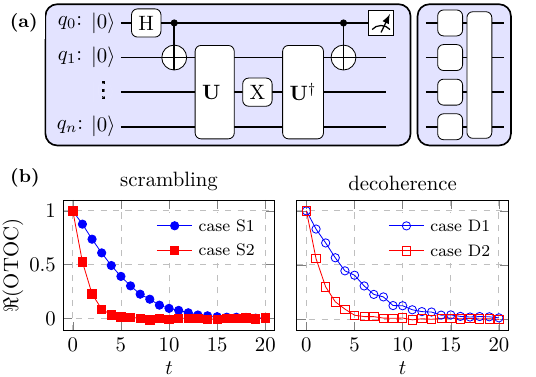}
    \caption{Measurements of the OTOC. (a) Quantum circuit for the interference protocol (left). Structure of a single evolution step of the fast scrambling model. (b) OTOC for the cases of scrambling and decoherence (Table~\ref{table}), which produces similar signals that can hardly be distinguished.} 
    \label{fig:otoc}
\end{figure}

As an illustration, we study a fast scrambling model recently proposed in Ref.~\cite{Belyansky2020Minimal}. The evolution unitary in the fast scrambling model consists of repetitive layers of circuit evolution. Each layer (Fig.~\ref{fig:otoc}a, right) is composed of Haar random single-qubit unitaries, immediately followed by a global entangling gate, i.e., which for $n$ qubits is given by
\begin{equation}
    \exp{\left(-i\frac{g}{2\sqrt{n}}\sum_{i<j}Z_iZ_j\right)},
\end{equation}
where $g$ is a constant parameter that controls the scrambling strength, and $Z_i$ is the Pauli $Z$ operator applied on the $i$th qubit. This building block can be viewed as a Trotterization of the evolution generated by a spin Hamiltonian with strong random local fields (generating single-gate random rotations) and all-to-all two-body couplings (creating the global entangling gate).

To simulate the effect of a noisy environment, we introduce errors in each layer of the single-qubit gates---after each single-qubit Haar random gate, a Pauli $X$ gate and a Pauli $Z$ gate are applied independently with probability $q$. Note that this error channel of the system qubits can be extended to a unitary evolution in an enlarged Hilbert space including ancillary qubits. Hence, the error model describes a decoherence process as well. We now have two parameters, $g$ and $q$, that control the strength of scrambling and decoherence, respectively. Table~\ref{table} lists the cases we considered. In the following numerical studies, we fix $n=10$.

\begin{table}[h]
\caption{\label{table}Cases compared for the fast scrambling model. The rate of scrambling increases with the value of $g$. The strength of the decoherence increase with the value of $q$.}
\begin{ruledtabular}
\begin{tabular}{ c c c c c c c } 
 Cases & S1 & S2 & D1 & D2 & I\\
 \hline
 $g$  & 1 & 2 & 0.5 & 0.5 & 1\\ 
 $q$  &  0 & 0 & 0.025 & 0.1 & 0.001 \\ 
\end{tabular}
\end{ruledtabular}
\end{table}

The OTOC measurement is achieved using the interference protocol developed in Ref.~\cite{Swingle2016Measuring}, as shown in the quantum circuit diagram in Fig.~\ref{fig:otoc}a. The intermediate $X$ gate is placed on the $i$th qubit $q_i$. The final measurement of the average value $\langle\sigma^0_x\rangle$ (or $\langle\sigma^0_y\rangle$) on the ancillary qubit $q_0$ then gives the real (or imaginary) part of the OTOC $\langle \sigma^0_x(t)\sigma^i_x\sigma^0_x(t)\sigma^i_x\rangle$, where $t$, in analog to time, is the number of layers in the forward (and hence the backward) evolution.

\begin{figure*}[t!]
    \centering
    \includegraphics[width=\textwidth]{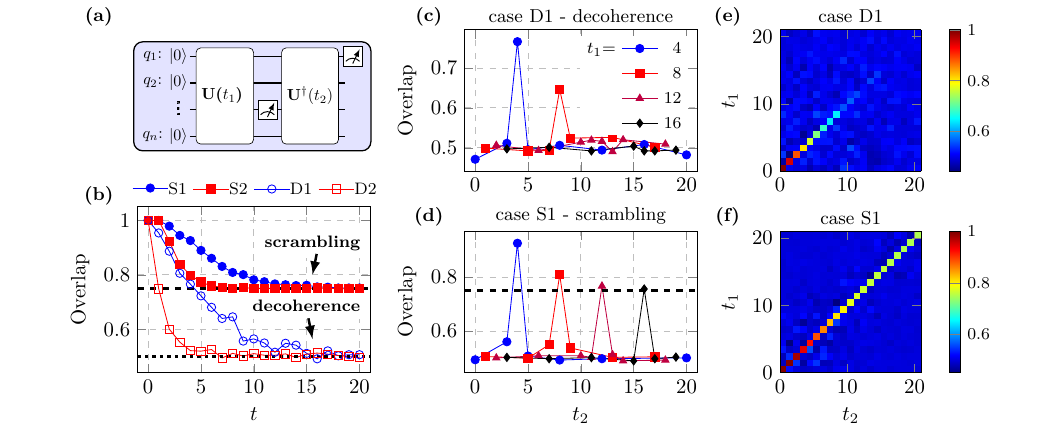}
    \caption{Measurement of the overlap $F(t)$ in (\ref{eq:fidelity}). (a) Schematic of the quantum circuit. (b) Decay of the overlap for the cases of scrambling and decoherence (Table~\ref{table}). (c,~d) The overlap as a function of the number of layers ($t_2$) in the backward evolution, for fixed numbers of layers $t_1$ in the forward evolution. (e,~g) Density plot for the overlap.} 
    \label{fig:recovery}
\end{figure*}

Figure~\ref{fig:otoc}b compares the evolution of the OTOC for ideal scrambling (without noise) and weak scrambling with strong decoherence, which exhibit similar decay curves. Hence, these two situations are practically indistinguishable from the OTOC measurements. 

We now examine the performance of our benchmark protocol for the fast scrambling model under the same conditions. The circuit diagram for this protocol is shown in Fig.~\ref{fig:recovery}a. Here, all the qubits are prepared in the computational $|0\rangle$. The recovery signal is obtained by measuring the overlap between the first qubit $q_1$ final state and its initial state. We also specify the intermediate perturbation between the forward and backward evolution as a projective measurement on a single qubit (other than the qubit 1) along the $Z$ axis. Note that since the evolution unitary contains Haar-random single-qubit gates in each layer, the measurement in a fixed direction is equivalent to random projective measurements. As discussed in the foregoing section, for this particular perturbation channel, the expected value of overlap (\ref{eq:fidelity}) is $0.75$ for ideal scrambling unitaries. On the other hand, in the case of strong decoherence, when the qubits eventually lose their coherence information, the overlap would be trivially $0.5$. Hence, the cases of scrambling with and without decoherence exhibit distinct asymptotic values of overlap. This is clearly demonstrated in Fig.~\ref{fig:recovery}b. 

To vitalize the emergence of the recovery signal produced by scrambling, we also performed simulations with different numbers of layers in the forward and backward evolution. That is, the overlap is measured after $t_1$ and $t_2$ layers of forward and backward evolution respectively, and if $t_2>t_1$, an additional $t_2-t_1$ layers in the backward evolution are chosen as independently random. Figures~\ref{fig:recovery}(c,~d) depict the overlap as a function of $t_2$ for fixed values of $t_1$. The density plots for the overlap scanned through various $t_1$ and $t_2$ are shown in Fig.~\ref{fig:recovery}(e,~f). The recovery signal emerges in a finite window around the peak $t_1=t_2$. The width of the peak reflects the timescale for local dissipation. For the current fast scrambling model, due to the random single-qubit rotation in every single layer, the recovery signal disappears as soon as $t_2$ is one layer away from $t_1$. It is worth noting that, in experiments with real time evolution, there can be a mismatch between the forward and backward evolution times $t_1$ and $t_2$, which contributes to the decay of the signal as well, together with decoherence caused by external couplings. Our protocol does not remove this source of errors, but positively confirms information scrambling when they are present.

The above simulations demonstrate that our protocol can unambiguously distinguish between scrambling and decoherence. In general, both of these two competing factors contribute to the decay of the overlap, but to different asymptotic values. This gives rise to a two-stage decay: In the early scrambling stage, the overlap decay is influenced by both scrambling and decoherence, until the information is fully scrambled and the overlap reaches the saturation value (\ref{eq:asymfidelity}). In the latter decoherence stage, the overlap further decays to a smaller value, at a rate determined purely by decoherence. The appearance of the two-stage decay indicates the presence of scrambling. 

\begin{figure}[h!]
    \centering
    \includegraphics[width=\columnwidth]{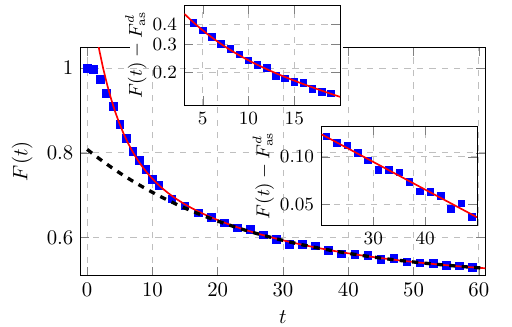}
    \caption{Decay of the overlap $F(t)$ for case I (Table~\ref{table}) with both scrambling and decoherence. Squares are numerical data. The red solid curve is the best fit to the ansatz (\ref{eq:ansatz}). To visualize the two stage decay of $F(t)$, we also plotted (black dashed) the long time pure exponential part of the ansatz (\ref{eq:ansatz}), which clearly departs from the early decay. Insets: $F(t)-F^d_{\rm as}$ in semilog scale, which show a single exponential decay at long times, and a sum of two exponential decays at early times. At the very beginning, $F(t)$ takes a quadratic form \cite{Yan2021Decoherence}.}
    \label{fig:benchmark}
\end{figure}

Suppose $\lambda_s$ and $\lambda_d$ are the exponential decay rates corresponding to scrambling and decoherence, respectively. In the extreme cases of strong scrambling and strong decoherence, the decay of $F(t)$ can be described as
\begin{equation}
  F(t) =
    \begin{cases}
      \left(1-F^s_{\rm as}\right) e^{-\lambda_s t} + F^s_{\rm as}, & \lambda_s \gg \lambda_d\\
     \left(1-F^d_{\rm as}\right) e^{-\lambda_d t} + F^d_{\rm as},  & \lambda_s \ll \lambda_d
    \end{cases}       
\end{equation}
where $F^s_{\rm as}$ and $F^d_{\rm as}$ are the asymptotic values of $F(t)$ induced by scrambling and decoherence, respectively. In the intermediate regime, such that the scrambling and decoherence are comparable, we propose an ansatz of $F(t)$:
\begin{equation}\label{eq:ansatz}
    F(t)=\left(a_1 e^{-\lambda_s t} + a_2 \right)e^{-\lambda_d t} + F^d_{\rm as}.
\end{equation}
With this, we can fit the observed overlap at long times (when the system gets sufficiently scrambled and the first exponential term vanishes) to a pure exponential function and extract the decoherence rate $\lambda_d$. Then the scrambling rate $\lambda_s$ can be extracted by fitting the early time decay to the ansatz (\ref{eq:ansatz}). We applied this procedure to case I (Table~\ref{table}), where both scrambling and decoherence contribute substantially to the decay of $F(t)$. Figure~\ref{fig:benchmark} shows simulations of $F(t)$, which is described accurately by ansatz (\ref{eq:ansatz}). This allows us to separate the scrambling rate $\lambda_s = 0.216$ and the decoherence rate $\lambda_d = 0.040$, and hence positively verify the presence of scrambling. The scrambling rate is determined purely by the underlying scrambling dynamics, and should not be altered by the intermediate perturbation. We verified this with different types of perturbations (\ref{eq:perturbation}) and recovery probabilities (\ref{eq:probasym}). The extracted scrambling rates are the same with a precision to the second decimal place. In Supplemental Material, we present simulations using an IBM cloud-based quantum computer. It clearly demonstrated that our protocol can successfully extract the correct scrambling rate under natural decoherence and gate errors of the current small size noisy quantum processors.

\vspace{5pt}
To summarize, we have developed a protocol to benchmark information scrambling, and examined it with both numerical studies and simulations on cloud-based quantum computers. This approach distinguishes between information scrambling and fake positive signals, produced by decoherence and operational errors in experiments unambiguously. It can be also used to quantify the degree of scrambling from the noisy background. Our method  requires only a single loop of forward and backward evolution, and hence can be applied to any system with access to time-reversing the dynamics.

\begin{acknowledgements}
This work was supported in part by the U.S. Department of Energy (DOE), Office of Science, Basic Energy Sciences, Materials Sciences and Engineering Division, Condensed Matter Theory Program, and in part by the U.S. DOE, Office of Science, Office of Advanced Scientific Computing Research, through the Quantum Internet to Accelerate Scientific Discovery Program. B.Y. also acknowledges support from the Center for Nonlinear Studies and the U.S DOE under the LDRD program in Los Alamos. J.H. was supported by the U.S. DOE through a quantum computing program sponsored by the Los Alamos National Laboratory (LANL) Information Science \& Technology Institute. This research used quantum computing resources provided by the LANL Institutional Computing Program, which is supported by the U.S. DOE National Nuclear Security Administration under Contract No. 89233218CNA000001. We also acknowledge the use of the IBM-Q LANL hub for NISQ device access in this work.
\end{acknowledgements}

\bibliography{reference}

\clearpage
\appendix

\setcounter{page}{1}
\renewcommand\thefigure{\thesection\arabic{figure}}
\setcounter{figure}{0} 

\onecolumngrid

\begin{center}
\large{ Supplemental Material for \\ ``Benchmarking Information Scrambling''
}
\end{center}

\section{Simulations on IBM quantum computers}

\subsection{The model}

This section presents an demonstration of the information scrambling benchmarking protocol on an IBM quantum processor. The model we studied here is a circuit model slightly different than the one presented in the main text. The total evolution consists of many $\emph{layers}$ of evolution, with each layer composed of two types of qubit operations. For instance, for $t$ layers, the evolution unitary is
\begin{equation}\label{eq:ut}
U_t = (U_2U_1)^t,    
\end{equation}
where $U_1$ is composed of single-qubit rotations, and $U_2$ is composed of two-qubit coupling gates.

\begin{figure}[h!]
    \centering
    \includegraphics[width=\textwidth]{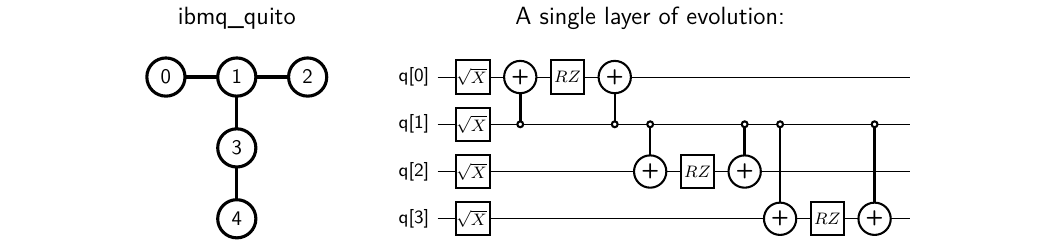}
        \label{fig:connect}
        \caption{Left: Qubit connectivity of the IBM-Q quantum processor ibmq\_quito. We use qubit labeled in $0$ to $3$ (q[0] - q[3]) as our system qubits. Right: The circuit diagram for a single layer of evolution, $U=U_2U_1$. $\sqrt{X}$ is the square root of the Pauli $X$ operator. $RZ$ is the rotation gate along the $Z$-axis for a fixed angle $g$ (fixed at $g=0.6$ in our simulations).}
\end{figure}

We will implement this model on IBM cloud-based quantum processors. In order to reduce the length of the circuit, we restrict the qubit operations to the native quantum gates, that is, gates that can be directly realized physically without conversion to compositions of other quantum gates. Hence, the single quantum gates are chosen as square root of the Pauli $X$ gate, i.e.,
\begin{equation}
        U_1 = \prod_i \sqrt{X_i}.
\end{equation}
The two-qubit gates are chosen as the $Z$-$Z$ coupling with a strength parameter $g$, i.e.,
\begin{equation}
        U_2 = \prod_{<i,j>} e^{-igZ_iZ_j/2}.
\end{equation}
However, the two-qubit couplings are not all-to-all, but restricted to the pairs of qubits $<i,j>$ that are directly connected in the quantum processors. For instance, Fig.~\ref{fig:connect} (left) shows the structure of connectivity of the IBM-Q quantum processor ibmq\_quito. (Couplings between pairs of qubits that are not naively connected on the hardware level can be realized as well, but at the cost of adding more quantum gates such as the SWAP gate.) We will use four qubits (q[0] - q[3]) as our system qubits to generate the scrambling unitary $U_t$. The circuit diagram of a single layer of evolution is plotted in the right panel of Fig.~\ref{fig:connect}.

\subsection{The simulation}

We implemented the proposed benchmarking protocol for scrambling as in{\rm tr}oduced in the main text. The simulation consists of the following steps: 1) Prepare the qubits (q[0] - q[3]) in a certain initial state. 2) Apply evolution $U_t$ (\ref{eq:ut}) with $t$ layers; Preform a projective measurement in the computational basis on qubit q[3]; Apply the reversed evolution $U_t^\dag$. 3) Measure the overlap, $F(t)$, between qubit q[1] final state and its initial state.

Since the system is relatively small, large fluctuations are expected in the measurement signal. To reduce such fluctuations, we performed several averaging strategies: The initial state of q[1] is prepared in the $+1$ eigenstate of the Pauli $X$ and $Y$ operator with equal probabilities. All the other qubits are initialized in a maximally mixed state. This is achieved by randomly sampling their initial states from $|0\rangle$ and $|1\rangle$ with equal probabilities. In the following, all the presented overlap signals are averaged with these procedures.

\begin{figure}[h!]
    \centering
    \includegraphics[width=\textwidth]{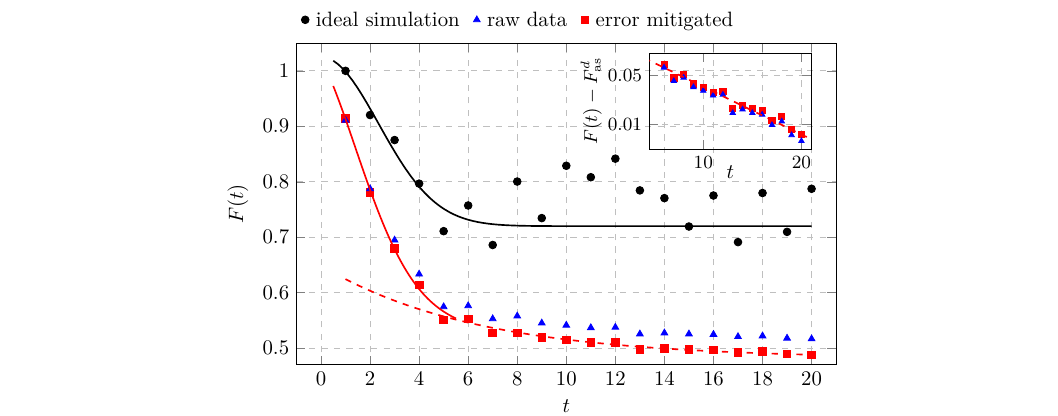}
        \label{fig:ibmq}
    \caption{Decay of the overlap for exact numerical simulations and measurements. The solid black curve is a Gaussian fit to the first 6 points of the numerical data. The dashed red curve is the exponential fit (\ref{eq:sm_exp}) to the late time decay of the error-mitigated data. The solid red curve is the Gaussian fit (\ref{eq:sm_Gau}) to the early time scrambling. Inset depicts the long time exponential decay (with the asymptotic constant term $F_{as}$ removed to reveal a clear linear curve in the semi-logarithm plot).}
\end{figure}

Figure~\ref{fig:ibmq} depicts the evolution of $F(t)$ of qubit q[1] as a function of $t$, the number of layers, up to $t=20$. First, exact numerical simulations show that $F(t)$ decays roughly to the universal value ($0.746$ for $4$ qubits) with a fluctuation caused by small size of the system. As discussed in the main text, $F(t)$ starts in a quadratic form. However, for small systems, the quadratic decay may switch to a Gaussian decay, before finally converting to an exponential decay \cite{Yan2020Information,Yan2021Decoherence}. This appears to be the case in the current study --- the first few data points of the numerical simulation fits very well to a Gaussian form with a decay (scrambling) rate $\lambda_{\rm s, exact} = 0.091$. Note that due to the strong finite size effect, the ex{\rm tr}acted value of the parameters are sensitive to the number of data points used in the fitting. We use the first $6$ points to ex{\rm tr}act the scrambling rate, and, to make the comparison fair, always use the same number of data points when fitting the measured data. The first $6$ points also give the best fit of the asymptotic value of $F(t)$ compared to the universal value.

We simulated the same process described above on the IBM-Q quantum processor ibmq\_quito, and performed measurement error mitigation to reduce measurement errors. As shown in Fig.~\ref{fig:ibmq}, the late time fluctuation of $F(t)$ is largely smeared out by strong decoherence, but still visible. The measured decay curves also exhibit clear late time exponential decay (see the inset of Fig.~\ref{fig:ibmq}),
\begin{equation}\label{eq:sm_exp}
    F(t) = ae^{-\lambda_dt}+F_{\rm as},
\end{equation}
with $\lambda_d$ fitted to $0.158$ (error mitigated data). This decay form is further used to fit the early scrambling regime to the ansatz 
\begin{equation}\label{eq:sm_Gau}
    F(t) = (be^{-\lambda_st^2}+c)e^{-\lambda_dt}+F_{\rm as},
\end{equation}
This allows us to ex{\rm tr}act the measured scrambling rates $\lambda_{\rm s} = 0.095$ (error mitigated data), which agrees with the exact value very well. This clearly demonstrates that our protocol can successfully extract the correct scrambling rate under natural decoherence and gate errors of the current small size noisy quantum processors.

\section{Universal form of the twirling channel}

This section collects a few basic formulas for the Bradbury process studied in the main text (also known as the quantum twirling channel in the quantum information literature) and sketches brief derivations of the universal form of the quantum states going through it.

For a generic quantum channel (which models the perturbation in the present context)
\begin{equation}
   \rho \rightarrow \Lambda(\rho) = \sum_k M^\dagger_k \rho M_k,\quad \sum_k M_kM_k^\dagger = \mathbb{I},
\end{equation}
the twirling of it is defined as
\begin{equation}
  \rho \rightarrow \Lambda_{\rm twirling}(\rho) = \int_{\rm Haar}dU\ U^\dag\Lambda(U \rho U^\dag)U,
\end{equation}
where the integral is performed over the unitary group with respect to the Haar measure. 

In general, Haar integral of higher moments of the unitaries can be computed with the aid of the Weingarten function \cite{Collins2003Monents}, which, for second moment, gives raise to the formula

\begin{equation}
\begin{aligned}
    \int_{\rm Haar} dU\ U_{m_1n_1}U^*_{m'_1n'_1}U_{m_2n_2}U^*_{m'_1n'_1}
    =& \frac{\delta_{m_1m'_1}\delta_{m_2m'_2}\delta_{n_1n'_1}\delta_{n_2n'_2}+\delta_{m_1m'_2}\delta_{m_2m'_1}\delta_{n_1n'_2}\delta_{n_2n'_1}}{d^2-1} \\
     & \quad\quad - \frac{\delta_{m_1m'_1}\delta_{m_2m'_2}\delta_{n_1n'_2}\delta_{n_2n'_1}+\delta_{m_1m'_2}\delta_{m_2m'_1}\delta_{n_1n'_1}\delta_{n_2n'_2}}{d(d^2-1)},
\end{aligned}
\end{equation}
where $d$ is the dimension of the unitary. This allows us to compute a general quantity

\begin{equation}\label{eq:sm_haarav}
\int_{\rm Haar} dU\ \ U^\dag A U B U^\dag C U
=\left[ \frac{{\rm tr}A\ {\rm tr}C}{d^2-1} - \frac{{\rm tr}(AC)}{d(d^2-1)}\right] B + \left[\frac{{\rm tr}(AC)\ {\rm tr}B}{d^2-1} - \frac{{\rm tr}A\ {\rm tr}C\ {\rm tr}B}{d(d^2-1)} \right] \mathbb{I}.
\end{equation}
Identifying $A=M^\dag_k$, $C=M_k$, and $B=\rho$, we get an explicit form of the twirling channel

\begin{equation}
\begin{aligned}
       \Lambda_{\rm twirling}(\rho) =&\sum_k \left[ \frac{{\rm tr}M^\dag_k\ {\rm tr}M_k}{d^2-1} - \frac{{\rm tr}(M^\dag_k M_k)}{d(d^2-1)}\right] \rho + \sum_k  \left[\frac{{\rm tr}(M^\dag_kM_k)\ {\rm tr}\rho}{d^2-1} - \frac{{\rm tr}M^\dag_k\ {\rm tr}M_k\ {\rm tr}\rho}{d(d^2-1)} \right] \mathbb{I}\\
       =& \frac{\sum_k {\rm tr}M^\dag_k\ {\rm tr}M_k -1 }{d^2-1} \rho + \left[\frac{d}{d^2-1} - \frac{\sum_k {\rm tr}M^\dag_k\ {\rm tr}M_k}{d(d^2-1)} \right] \mathbb{I}\\
       =& p\rho + (1-p)\frac{\mathbb{I}}{d},
\end{aligned}
\end{equation}
with probability
\begin{equation}
    p = \frac{\sum_k {\rm tr}M^\dag_k\ {\rm tr}M_k -1 }{d^2-1}.
\end{equation}

As an example, consider $\Lambda$ a single-qubit projective measurement on a $n$-qubit system with total dimension $d$. The Kraus operators for $\Lambda$ read $M_0 = |0\rangle\langle 0|$ and $M_1 = |1\rangle\langle 1|$. The total system is prepared in an initial product state $|\phi\rangle\langle\phi|\otimes\rho$, where $|\phi\rangle$ is a pure state of the single qubit that the projective measurement applies on, and $\rho$ is the initial state of the rest of the total system. In this case, the output state through the twirling channel becomes
\begin{equation}
   \Lambda_{\rm twirling}\left(|\phi\rangle\langle\phi|\otimes\rho\right) =  p|\phi\rangle\langle\phi|\otimes\rho + (1-p)\frac{\mathbb{I}}{d},
\end{equation}
with
\begin{equation}
    p = \frac{d^2/2 -1 }{d^2-1}\overset{d\rightarrow\infty}{\longrightarrow} 0.5.
\end{equation}
The overlap between the single qubit output state and its initial state, is
\begin{equation}
    F = {\rm tr}\ \left(|\phi\rangle\langle\phi|\otimes\mathbb{I}\right)\Lambda_{\rm twirling}(|\phi\rangle\langle\phi|\otimes\rho) = (1+p)/2 \overset{d\rightarrow\infty}{\longrightarrow} 0.75.
\end{equation}
In contract, the overlap behaves generally differently in the presence of decoherence. For example, for strong decoherence that is described by a depolarizing channel, asymptotically, the output state through the same twirling channel will become a totally mixed state. Hence, the overlap becomes
\begin{equation}
    F = {\rm tr}\ \left(|\phi\rangle\langle\phi|\otimes\mathbb{I}\right)/d = 0.5.
\end{equation}
Distinction between these two different values of overlap makes it possible to distinguish between scrambling and decoherence. On the other hand, the OTOC of local observables typically exhibits the same asymptotic values with and without decoherence~\cite{Yoshida2019Disentangling}.

\end{document}